\documentclass[english]{article}
\usepackage[T1]{fontenc}
\usepackage[latin9]{inputenc}
\usepackage{graphicx}
\usepackage{babel}

\begin{document}

\title{Marking Systemic Portfolio Risk with Application to the Correlation
Skew of Equity Baskets}

\author{Alex Langnau\\
 Daniel Cangemi\\
}

\maketitle
The downside risk of a portfolio of (equity)assets is generally substantially
higher than the downside risk of its components. In particular in
times of crises when assets tend to have high correlation, the understanding
of this difference can be crucial in managing systemic risk of a portfolio.
In this paper we generalize Merton\textquoteright{}s option formula
in the presence jumps to the multi-asset case. It is shown how common
jumps across assets provide an intuitive and powerful tool to describe
systemic risk that is consistent with data. The methodology provides
a new way to mark and risk-manage systemic risk of portfolios in a
systematic way.

\pagebreak{}

\part*{Introduction}

It has been argued that one of the factors that triggered the downfall
of Long Term Capital Management (LTCM) was its failure to incorporate
\textquotedblleft{}fat tails\textquotedblright{} of asset price distributions
properly into investment decisions as well as risk management {[}1{]}.
Today financial institutions systematically deduce fat tails from
the option markets and incorporate this information consistently into
the pricing as well as the risk management framework. Despite this
progress it is interesting to note that little or even no efforts
have been made in the marking and risk-managing of \textquotedblleft{}fat
tails\textquotedblright{} when individual assets are \emph{combined}
into a large portfolio. However a large part of a portfolio\textquoteright{}s
fat tail is not the result of the fat tails of the individual components.
It is this difference that can be associated with systemic risk of
a portfolio that will generally dominate investment decisions in times
of crises.\\
\\
As pointed out {[}2{]} only roughly 50\% of the downside risk
(fat tail) of a portfolio of stocks such as the DAX, is due to the
fat tails of its members. In a sense by ignoring the additional 50\%
of downside, financial markets have not yet fully learned the lessons
from the LTCM debacle. In fact the current financial crisis is haunting
us with the consequences of underestimating systemic risk in portfolio
management.\\
\\
In this paper we argue that financial institutions should extend
their market data to account for systemic risk not only in the area
of credit but across all assets classes whenever assets are combined
into a portfolio. The proposed \textquotedblleft{}extension of market
data\textquotedblright{} would be similar, in some respect, to the
generalization of at-the-money volatilities to all strikes.\\
\\
The first objective of this paper is to show that the steepness
of the DAX skew compared to its components can be linked to the systemic
risk of the \textquotedblleft{}DAX portfolio\textquotedblright{}.
This is important because it allows us to infer the systemic risk
from the market consistently. The second objective of this paper is
to explicitly construct a parameterization that describes the systemic
risk of large equity portfolios. The latter can serve as a candidate
for the extension of the market data mentioned above.\\
\\
Recently, efforts have been made to extend Dupire\textquoteright{}s
local volatility framework to incorporate state-dependent correlations
into the dynamics of a portfolio {[}2, 3{]}. While this leads to a
numerically efficient generalization of Dupire\textquoteright{}s model,
there are two main drawbacks. \\
\\
First, the model requires an a priori knowledge of the basket
skew as an input. This is however not known in many cases. The method
presented in this paper provides a natural way to interpolate skews
to sub-baskets as well as cross-index baskets. \\
\\
Second, while the extension of Dupire\textquoteright{}s model
to local correlation consistently describes the correlation skew,
it does not attempt to \textquotedblleft{}explain\textquotedblright{}
its causes. Consequently, it cannot be viewed as a straightforward
methodology to monitor systemic risk. The approach presented in this
paper defines a copula that explicitly links the correlation skew
to systemic risk and hence could be a candidate to improve the (tail)
risk-management of large portfolios. Work in this direction has been
pursued by others {[}4{]},{[}8{]}. 
 The paper is organized as follows: The
first section reviews Merton\textquoteright{}s option formula that
describes options in the presence of jumps. We then generalize this
formula to the multi-asset case and show qualitatively how correlation
skew can be induced in this framework. The next chapter generalizes
the single-asset Merton formula in order to properly reproduce market
prices of options at all strike prices. Finally we put both generalizations
together and define the \textquotedblleft{}Merton-Copula basket\textquotedblright{}
that enables one to define systemic portfolio risk in a proper way.

\part*{A Merton basket option formula}

In 1976, Merton generalized the Black-Scholes formula to situations
where jumps are present {[}5{]}. In this case the dynamics of the
stock price $S_{t}$ is given by

\begin{equation}
\frac{dS_{t}}{S_{t}}=\left(r_{t}-q_{t}-\lambda E_{\lambda}\left[Y-1\right]\right)dt+\left(Y-1\right)dn_{\lambda}+\sigma dw_{t}\end{equation}
where $r_{t},q_{t},\lambda,dn_{\lambda},\sigma,w_{t}$ denote the
short rate, dividend yield, jump intensity, jump measure, diffusive
vol and a Brownian motion respectively. The jump size $Y-1$ has a
mean $\hat{k}$ and a lognormal volatility $+\delta$, i.e.\[
Y=\left(1+\hat{k}\right)e^{-\frac{1}{2}\delta^{2}+\delta z}\]
where $z\in N\left(0,1\right)$. Equation 1 states that the dynamics
of $S_{t}$ is log-normal with occasional jumps occurring at rate
$\lambda$ and with size $Y-1$.\\
\\
As the jumps are assumed proportional (rather than additive shifts)
as well as fully uncorrelated to the diffusive part of the dynamics,
the Merton price of a call option Call(T,K) with maturity T and strike
K is simply given by an infinite sum of Black Scholes prices, i.e.\begin{equation}
Call\left(T,K\right)=\sum_{n=0}^{\infty}e^{-\lambda T}\frac{\left(\lambda T\right)^{n}}{n!}BS\left(F\left(1+\hat{k}\right)^{n}e^{-\lambda \hat{k} T},T,K,\sqrt{\sigma^{2}+\frac{n\delta^{2}}{T}}\right)\end{equation}
 where\begin{equation}
BS\left(F,T,K,\sigma\right)=Df\left(F\, N\left(d_{1}\right)-K\, N\left(d_{2}\right)\right)\,:\, d_{1/2}=\frac{log\left(\frac{F}{K}\right)\pm\frac{1}{2}\sigma^{2}T}{\sigma\sqrt{T}}\end{equation}
 denotes the standard Black-Scholes option formula.\\
\\
It is very easy to interpret Eq.2 in intuitive terms: $e^{-\lambda T}\frac{\left(\lambda T\right)^{n}}{n!}$
is the probability of n jumps happening until maturity. In this case
the spot has suffered an extra depreciation of $\left(1+\hat{k}\right)^{n}$
on average due to the jumps which is reflected in the adjustment to
the forward price in the Black-Scholes formula.\\
\\
The Merton model describes a simple way to incorporate volatility
skew into the dynamics without loss of analytic tractability. However
the model is too simple to give an exact fit to the observed volatility
skew in practice. In Fig.1 we plot the implied volatility skew for
different levels of the diffusive volatility $\sigma.$ As $\sigma$
increases the jump generates less skew which is consistent with intuition.\\
\\
\\
\includegraphics[scale=0.75]{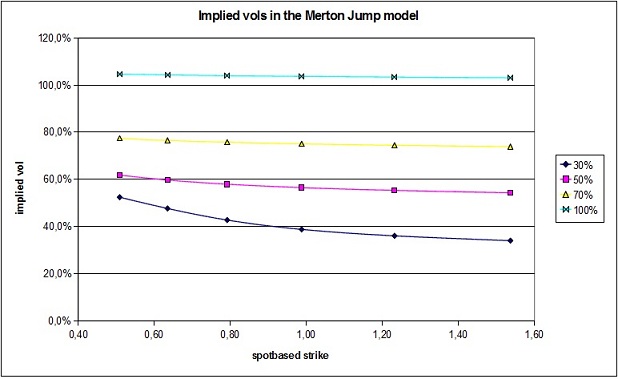}\\
\\

Fig.1: The volatility skew for different levels of diffusive vol.
The effect of jumps on the skew becomes less and less pronounced for
high values of the diffusive vol.\\
\\
In the Appendix 1 we show that a way to retain a volatility skew for large $\sigma$ stocks is
to scale the jump size with the diffusive vol itself, e.g.

\begin{equation}
\hat{k}\rightarrow\left(\frac{\sigma}{\sigma_{0}}\right)^{\kappa}\hat{k}\end{equation}
 $\sigma_{0},\kappa$ denotes a \textquotedblleft{}vol-scale\textquotedblright{}
and \textquotedblleft{}vol-scale elasticity\textquotedblright{} respectively.\\
\\
In order to generalize Merton\textquoteright{}s formula to basket
options, Eq.1 can be generalized to N assets that evolve according
to\\
\\
\begin{eqnarray}
\frac{dS_{t}^{\left(i\right)}}{S_{t}^{\left(i\right)}}\hspace{0.5in} & = & \left(r_{t}-q_{t}^{\left(i\right)}-\lambda 
E_{\lambda}\left[Y^{\left(i\right)}-1\right]dn_{\lambda}\right)+\sigma^{\left(i\right)}dw_{t}^{\left(i\right)}+\left(Y^{\left(i\right)}-1\right)dn_{\lambda},\, \nonumber \\
\left\langle dw_{t}^{\left(i\right)},dw_{t}^{\left(j\right)}\right\rangle  & = & \rho_{ij}^{diffusive}dt\, \hspace{0.5in}\, \,i=1,\ldots,N\end{eqnarray}
 Note that, despite the multi-assets character of Eq.5 only a single
Poisson process is \textquotedblleft{}employed\textquotedblright{}
just like in the case of equation 1. As we show later, this fact will
prove crucial in order to generate correlation skew as well systemic
risk of a portfolio.\\
\\
Eq.5 states that the dynamics between jump events is governed
by a multi-asset Black-Scholes dynamics with a diffusive correlation
$\rho_{ij}^{diffusive}$\\
\\
In the following we define a basket\begin{equation}
B_{t}=\sum_{i=1}^{n}\alpha_{i}S_{t}^{\left(i\right)}\end{equation}
 with fixed basket weights $\alpha_{i}$. If Q denotes the pricing
measure, Df the discount factor and $\hat{k}_{i}$ the jump size of
asset i, then a call option on a basket can be calculated according
to\begin{equation}
\left(Basket\right)-Call\left(T,K\right)=Df\, E^{Q}\left(\left(B_{t}-K\right)^{+}\right)\end{equation}
 \[
=\sum_{n=0}^{\infty}e^{-\lambda T}\frac{\left(\lambda T\right)^{n}}{n!}TMPricer\left(F_{1}\left(1+\hat{k}_{1}\right)^{n}e^{-\lambda \hat{k_{1}}T},\ldots,F_{N}\left(1+\hat{k}_{N}\right)^{n}e^{-\lambda\hat{k}_{N}T},\rho_{ij}^{diffusive}\right)\]
 \\
Note that contingent to the number of jumps, the dynamics is log-normal
and can hence well be approximated by a well known Black-Scholes three-moment
basket pricer denoted by $TMPricer\left(F_{1},\ldots,F_{N},\rho\right)$(
see for example {[}6{]} ).\\
\\
Equation 7 can be viewed as the generalization of Merton\textquoteright{}s
formula to options on a basket of assets.\\
\\
In Eq.5,7 all assets \textquotedblleft{}share\textquotedblright{}
the same Poisson process defined by $\left(\lambda,\hat{k},\delta\right).$
This triple defines a Poisson process with intensity $\lambda$, a
universal jump size $\hat{k}$ and jump volatility $\delta$.\\
\\
Individual jump-sizes $\hat{k}_{i}$ are computed from the \textquotedblleft{}universal
jump-size\textquotedblright{} by means of \begin{equation}
\hat{k}_{i}=\left(\frac{\sigma_{i}}{\sigma_{0}}\right)^{\kappa}\hat{k}\hspace{1in}i=1,\ldots,N\end{equation}
 In the following we present results of Eq.7 in combination with Eq.8
when applied to the components of the DAX. Fig.2 shows the Merton
basket implied vols that are inferred from Eq.7 versus the vol of
the DAX for $\left(\lambda,\hat{k},\delta,\sigma_{0}\right)=\left(25\%,-16\%,18\%,18\%\right)$\\
Table 1 shows how parameters change with the time horizon under consideration. 
\\

\includegraphics[scale=0.95]{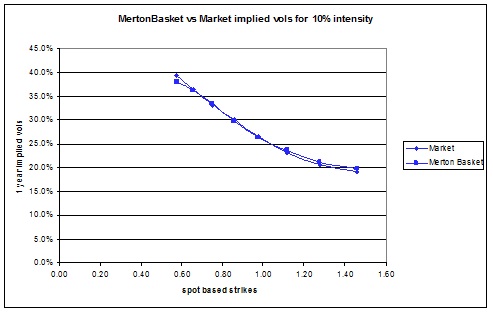}

Fig.2: Reconstruction of DAX skew from its components: Merton basket
implied vols versus market for $\left(\lambda,\hat{k},\delta,\sigma_{0}\right)=\left(25\%,-16\%,18\%,18\%\right)$\\
\\
\\

\begin{center}
\begin{scriptsize}

\label{decomp}

\begin{tabular}{|rr|rr|rr|rr|rr|}
\hline
\hline
Maturity  & $\lambda$ & $\hat{k}$ & $\delta{k}$ & $\sigma_{0}$& \\
\hline
1 Year	&0.25&-0.13	&18\%&0.12	&\\
2 Year	&0.25&-0.137	&15\%&0.12 &\\
3 Year	&0.25&-0.13	&18\%&0.127	&\\
\hline
\hline
\end{tabular}
\end{scriptsize}
\end{center}

Table 1: Jump parameter dependence as a function of time that is required to reconcile the skew of the DAX from its components. The parameters appear high relative to historic values. However they, in parts, reflect risk-premia that are associated with the sale of deep out-of-the-money put-options.
 \vspace{5mm}

It is remarkable that the simple formulas of Eq.7,8 leads to quite
good agreement with the skew of the DAX. We remind the reader that
typically only 50\% percent of the DAX skew can be explained in terms
of the skew of the underlying assets {[}2{]}. Hence the idea of \textquotedblleft{}sharing\textquotedblright{}
a Poisson process across different assets seems an idea worth investigating
as a means to create copulas that are consistent with data. The intuition
behind this concept is easily explained: When a far out-of-the money
put on a basket ends up in the money at time of maturity, it is likely
that at least one jump has occurred during the life of the option.
As jumps occur simultaneously across all assets in our model, they
effectively increase the implied correlation for out-of-the-money
puts giving rise to correlation skew. The situation is reversed for
out-of-the money call options that are likely to pay out only when
no jump event has occurred. The implied correlation is expected to
be close to the diffusive correlation in this case. For $\delta=0$
the relationship between the diffusive correlation $\rho_{ij}^{diffusive}$
and total expected correlation $\rho_{ij}$ can be calculated explicitly
with the result\begin{equation}
\rho_{ij}=\frac{\rho_{ij}^{diffusive}+\lambda\hat{k}_{i}\hat{k}_{j}}{\sqrt{\left(1+\lambda\hat{k}_{i}^{2}\right)\left(1+\lambda\hat{k}_{j}^{2}\right)}}\end{equation}
 Note that in the limit of very frequent jumps where $\lambda\rightarrow\infty$
the correlation goes to one confirming our intuition. Eq. 9 plays
a useful role when calibrating the set $\left(\lambda,\hat{k},\delta,\sigma_{0}\right)$
to the skew of the DAX. This is because $\rho_{ij}$ needs to be kept
constant in order not to change the levels of the at-the-money basket
vols for different choices of $\lambda$ during calibration. Higher
levels for the intensity require lower diffusive correlations which
can be deduced from Eq.9.\\
\\
Even though this explanation is highly plausible, it is not the only mechanism that yields tp correlation skew. For example, multi-asset stochastic volatility models that exhibit negative cross-correlations between spot returns of one asset with the vol of another asset generate cross-vanna sensitivity and hence correlations skew. It would be  interesting to combine this effect with the one outlined above to a more complex model. The latter would decrease the intensities as well as the jumps-sizes of Table 1 that are required to achieve consistency between the index skew and the skew of its components. In many respects such a model could be considered the next step in the devlopment that is somewhat similar to progress in the modelling of FX derivatives where two different mechanism for the volatility skew- local vol and stochastic vol get combined to a more realistic description of this smarket.
\\

We summarize this chapter by stating that the ideas presented
seem worthwhile perusing. However the Merton models as well as the
Merton basket formula need to be extended in order to make the calibration
to the individual stock skews exact.

\newpage{}

\part*{Single asset Merton extension}

The main drawback of Merton\textquoteright{}s model is its failure
to calibrate option prices exactly to the market. An example for the
mis-calibration is given in Fig.3.\\
\\
\includegraphics[scale=0.85]{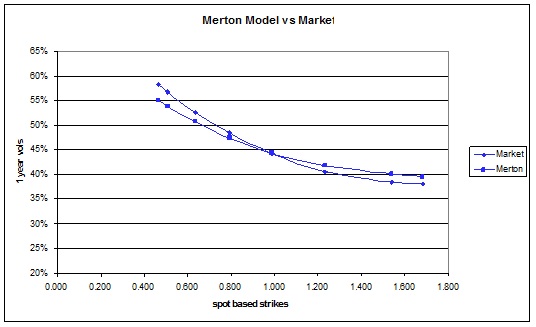}

Fig.3:Example for mis-calibration of a single stock skew in the Merton
model.\\
\\
In order to fix this problem we generalize the Merton Single asset
formula of Eq.2 to\[
Call\left(T,K\right)=\]
\begin{equation}
\sum_{n=0}^{\infty}e^{-\lambda T}\frac{\left(\lambda T\right)^{n}}{n!}\, BS\left(F\left(1+\lambda \hat{k}\right)^{n}e^{-\lambda \hat{k}T},T,K,\sqrt{\sigma\left(K\right)^{2}+\frac{n\delta^{2}}{T}}\right)\end{equation}
 The only difference is that the diffusive vol itself becomes a function
of the strike price. The idea is that for a given set $\left(\lambda,\hat{k},\delta\right)$
the calibration mismatch is picked up by $\sigma\left(K\right)$.\\
\\
The function $\sigma\left(K\right)$ can easily be deternimed
by means of the fixed point algorithm {[}6{]}: The The algorithm starts
with a constant zeroth guess $\sigma$ for the diffusive volsurface
$\sigma^{\left(0\right)}\left(K_{j}\right)=\sigma$ for M different
values of strike $K_{j}$, $j=1,\ldots,M$. Iteration $i+1$ is obtained
from iteration $i$ according to $\sigma^{\left(i+1\right)}(K_{j})=\sigma^{\left(i\right)}\left(K_{j}\right)+\Delta\sigma_{j}$where
$\Delta\sigma_{j}$ is the vol-mismatch between the implied vol inferred
from Eq.10 using $\sigma^{\left(i\right)}\left(K_{j}\right)$ and
the market implied vol. For reasonable first guesses the algorithm
typically converges very fast as the following diagrams show\\
\\
\includegraphics{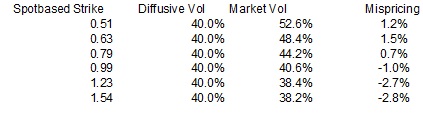}\\
\includegraphics{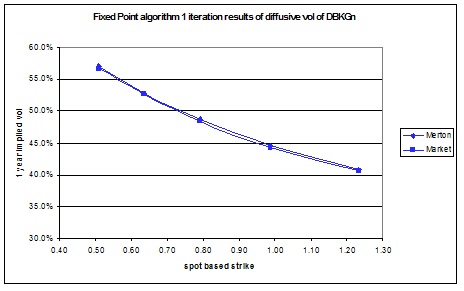}

Fig.4: First iteration results for the fixed point algorithm:\\
\\
Second iteration results for the fixed point algorithm:\\
\\
\includegraphics{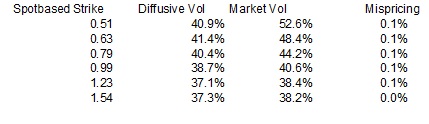}\\
\includegraphics{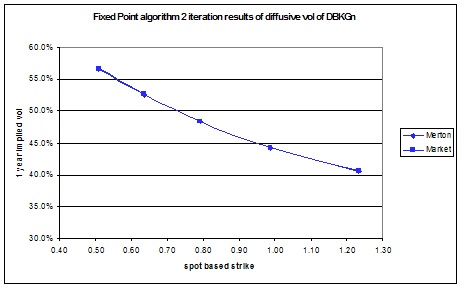}\\
Fig.5: Second iteration results for the fixed point algorithm:\\
\\
Equation 10 together with the fixed-point algorithm for the determination
of the strike dependent diffusive skew defines the our \textquotedblleft{}Merton
single asset extension formula\textquotedblright{} for the pricing
of options in the presents of jumps as well as \textquotedblleft{}diffusive\textquotedblright{}
contributions to the skew.\\
\\
It is important to realize that one can easily construct a stochastic
process explicitly that yields Eq.10 in the presence of \textquotedblleft{}diffusive
skew\textquotedblright{} $\sigma(K)$ for example:\begin{equation}
\frac{dS_{t}}{S_{t}}=\left(r_{t}-q_{t}-\lambda E_{\lambda}\left[Y-1\right]\right)dt+\left(Y-1\right)dn_{\lambda}+\hat{\sigma}\left(X_{t},t\right)dw_{t}\end{equation}
 where $\hat{\sigma}\left(X_{t},t\right)$ can be thought of a some
kind of \textquotedblleft{}local vol\textquotedblright{} with the
difference that it is not a function of the spot $S_{t}$ but rather
the variable $X_{t}=\frac{S_{t}}{Y^{n}}$ where n is the number of
jumps that have occurred up to time t.

\newpage{}

\part*{The Merton Copula Basket}

The goal of this chapter is to incorporate the results of the previous
chapter into a basket options formula along the lines of Eq.7. However
this is slightly more difficult as the three-moment approximation
that was applied in Eq.7 is not appropriate in this case. This is
because contingent to a specified number of jumps the dynamics is
not log-normal anymore but given by a local volatility of Eq.11. In
the following we outline the generalization of the three-moment pricer
that was implicit in Eq.7 to \textquotedblleft{}infinite\textquotedblright{}
moments denoted by $BasketCopulaPricer\left(S_{i},K,\rho,\sigma_{i}(K)\right)$
\begin{itemize}
\item \textbf{Step 1}: Dial a set of correlated normal deviates $w_{i}\, i=1,\ldots,N$
from a correlation matrix $\rho_{ij}$ and convert them into uniform
deviates according to $u_{j}=N\left(w_{j}\right)$
\item \textbf{Step 2}: Invert the individual asset (option implied) distributions
according to\[
u_{i}=1+\frac{d}{dK}\frac{1}{Df}call\left(T,K\right)\]
 to obtain a set $\left(S_{T}^{\left(1\right)}\left(u_{1}\right),S_{T}^{\left(2\right)}\left(u_{2}\right),\ldots,S_{T}^{\left(N\right)}\left(u_{N}\right)\right)$
\item \textbf{Step 3}: Calculate $payoff=\left(\sum\limits _{i=1}^{N}\alpha_{i}S_{t}^{\left(i\right)}-K\right)^{+}$
\item \textbf{Step 4}: Go to step 1 Npath times, then average results and
obtain
\end{itemize}
\[
Call\left(T,K\right)\equiv BasketCopulaPricer\left(S_{i},K,\rho,\sigma_{i}(K)\right)\]
 \\
We are now in a position to define the Merton Copula basket price
in the presence of jumps as well as \textquotedblleft{}local volatility\textquotedblright{}
(see Eq.11)\\
\begin{equation}
Call\left(T,K\right)=\sum_{n=0}^{\infty}e^{-\lambda T}\frac{\left(\lambda T\right)^{n}}{n!}BasketCopulaPricer\left(S_{i}\left(1+\hat{k}_{i}\right)^{n}e^{-\hat{k}_{i}T},K,\rho_{ij}^{diffusive},\sigma_{i}(K)\right)\end{equation}
 where $\sigma_{i}(K)$ is given by the solution of the fixed point
algorithm.\\
\\
Fig.6 shows the results in the case of $\lambda=0$ recovering
the regular Gaussian Copula result. As mentioned in the beginning
of the paper that not even 50\% percent of the skew of the basket
is recovered in this case from the skew of the individual assets.\\
\\
\includegraphics[scale=0.75]{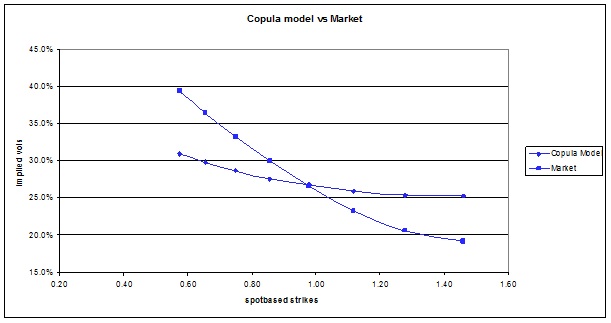}

Fig.6: The Merton Copula basket reduces to the standard Gaussian Copula
in the case of $\lambda=0$. Less than 50\% of the skew of the basket
is recovered in this case.\\
\\
Fig.7 shows the results after calibration. Note that for all choices
of the calibration set $\left(\lambda,\hat{k},\delta,\kappa\right)$
the approach is consistent with the skews of the individual assets.\\
\\
\includegraphics[scale=0.77]{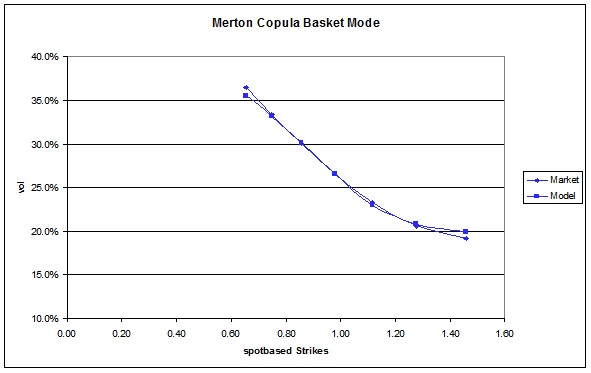}

Fig.7: Calibration results of the DAX skew from its components for
the set $\left(\lambda,\hat{k},\delta,\kappa\right)=\left(0.35,-0.16,18\%,1,17,18\%\right).$\\
\\
Hence the set $\left(\lambda,\hat{k},\delta,\kappa\right)$ describes
\textquotedblleft{}orthogonal new market-data\textquotedblright{}
that quantifies the correlations skew. One expects a different set
for each maturity.

\newpage{}

\part*{Summary}

We presented a generalization of the Merton Jump Option formula to
multi-assets as well as diffusive skews and found that common jumps
are an appropriate and intuitive way to define copulas that describe
the systemic risk of a portfolio. Financial institutions should start
treating systemic risk as \textquotedblleft{}market data\textquotedblright{}
across asset classes and start to mark, risk-manage and price these
effects in a systematic way. For each maturity, a tuple $\left(\lambda,\hat{k},\delta,\kappa\right)$
can serve as a candidate to describe this systemic risk in an appropriate
way. It is straightforward to generalize our framework to more than
one Poisson jump in the case different sources of systemic risk are
encountered. This makes the framework well suited to mark skews of
sub-baskets as well as cross-index baskets.\\
\\

\part*{Appendix 1}

In this Appendix we motivate the scaling of the jumpsize by the volatility $\sigma $ according to Eq.4 for $\kappa = 1$ in order to retain the volatility skew for large  $\sigma $.
 
From Eq.2 the price of a Digital-Put option of strike K and maturity T $ Digital-Put(K,T)= E^{\beta}\left(1_{S_{T}<K}\right) $ is given by

\begin{equation}
Digital-Put\left(T,K\right)=\sum_{n=0}^{\infty}e^{-\lambda T}\frac{\left(\lambda T\right)^{n}}{n!}N(-d_{2})\end{equation}

Hence the price change of the digital with a change in intensity $\lambda $ is given by

\begin{equation}
\frac{\partial Digital-Put\left(T,K\right)}{\partial \lambda}=\sum_{n=0}^{\infty}e^{-\lambda T}\frac{\left(\lambda T\right)^{n}}{n!} 
\varphi(-d_{2})  \frac{ln \frac{ F(1+k)^{n} }{K}}{\sqrt{T}} \frac{k^2}{\sigma^2} 
\end{equation}

where $\varphi(x)$ denotes the normal density.This is due to the fact that for small values for $\lambda$ the relationship between implied and diffusive vol can be approximated by
$\sigma^2(implied)=\sigma^2(diffusive)+\lambda k^2$ and the last term in the numerator of Eq.3 is generally small for suffently out-of-the money options.
Hence if one wants to retain the skew for large values of $\sigma$, the factor in Eq.14 suggests to scale the jump-size according to $k->\sigma k$.



\part*{References}
\begin{enumerate}
\item When Geniuses Failed: The Rise and Fall of Long-Term Capital, ISBN
0-375-75825-9
\item A dynamic model for correlation, RISK magazine Mar 2010, Alex Langnau
\item When Correlation breaks, RISK magazine Sep 2010, Adil Reghai
\item Alexander Lipton, Arthur Sepp, Credit value adjustment for credit
default swaps via the structural default model, The Journal of Credit
Risk (2009)
\item Continuous-Time Finance, R.C. Merton, Blackwell 1990
\item The hybrid most likely path, RISK April, Adil Reghai
\item Approximated moment-matching dynamics for basket options simulations,
Brigo,Mercurio, Rapisarda, Scotti, 2001

\item Equity Correlations Implied by Index Options: Estimation and Model Uncertainty Analysis
Rama Cont,Romain Dequest, April 19, 2010

\end{enumerate}

\part*{Acknowledgements}

One of us (A.L.) would like to thank Brian Norsk Huge for constructive
comments during the European Quant congress 2010.
\end{document}